\documentclass[runningheads]{llncs}
\usepackage{graphicx}
\usepackage{tikz}
\usetikzlibrary{decorations.pathmorphing} 
\usetikzlibrary{fit}					
\usetikzlibrary{backgrounds}	
\usetikzlibrary{shapes,arrows}
\usepackage{xcolor} 

\begin{document}
\title{Election in India: Polling in National Financial Switch}
%
%
\author{Subhankar Mishra\inst{1,2}\orcidID{0000-0002-9910-7291} 
}
\authorrunning{S. Mishra}
%
\institute{School of Computer Sciences, 
National Institute of Science Education and Research, 
Bhubaneswar, Odisha 752050, India \and
Homi Bhabha National Institute.
Anushaktinagar, Mumbai, 400094
India
\email{smishra@niser.com}\\
\url{https://www.niser.ac.in/users/smishra} 
}
%


\maketitle              
\begin{abstract}
Indian voters from Kashmir to Kanyakumari select their representatives to form their parliament by going to polls. India's election is one of the largest democratic exercise in the world history. About 850 million eligible voters determine which political party or alliance will form the government and in turn, will serve as prime minister. Given the electoral rules of placing a polling place within 2 kilometers of every habitation, it comes as no surprise that is indeed a humongous task for the Election Commission of India (ECI). It sends around 11 million election workers through tough terrains to reach the last mile. This exercise also comes as ever growing expenditure for the ECI. This paper proposes the use of Automated Teller Machines (ATM) and Point Of Sale (POS) machines to be used to cover as much as urban, rural and semi-urban places possible given the wide network of National Financial Switch (NFS) and increase in connectivity through Digital India initiative. This would add to the use of the existing infrastructure to accomodate a free, fair and transparent election.

\end{abstract}

\section{Introduction}

Elections in India have become the most visible symbol of democratic process today. In a large democracy such as India, there is necessity of representation as all the citizens cannot directly take part in decision making. This marks the importance of the elections. When a decision has to be taken by hundreds of crores of people, direct democracy cannot be practiced. Hence the rule by the people is mostly rule by people's representatives across the world and also in India. After the results of the election are declared, the leading party is asked to form the government by the President of India. If there is no single party with an absolute majority, a coalition is formed to reach the majority. There are 543 directly elected members of parliaments (MPs). 

Election in India is achieved through voting on Electronic Voting Machines. While they have been paraded with the fairness by the ECI, it has had its own share of stories that have totally questioned credibility. Security of EVMs has been debated a lot in scientific literature \cite{evm2010}. There have been also attempts at using the newer technologies such as block chain for e-voting \cite{Osgood2016,Kshetri2018}, where they do correlate the blockchain enabled cryptocurrency techniques to enable wallet system for voting, where voter could use his/her coin only once for voting in a given election. There has also has been numerous patents \cite{Patent2003,Patent2005,Patent2008,Patent2011,Patent2018} by Google on using the Automated Teller Machines as voting machines and using the secure architecture already built by the same for polling.

However, even though India has a good network of ATMs across the country, its reach is only limited to urban and semi-urban areas. However, rural areas are still devoid of ATMs. Hence the proposal is also to include the POS machines to reach the last mile and connect to remote places. Although the proposal would still need the use of the electronic voting machines where it is difficult for the establishing connectivity through ATMs or even POS machines. 

Section \ref{election} describes some facts about the elections in India. 

\section{Elections in India} \label{election}
Elections in India happens through people voting physically at different polling stations through Electronic Voting Machines (EVM). 

\subsection{Voting}
Elections are held in Parliamentary constituencies across States and Union Territories in India. Some of the facts of the general election 2019 are: 
\begin{itemize}
    \item 2019 general election was held for 543 Parliamentary constituencies across 29 States and 7 Union Territories.
    \item 900 million people, above 18 years of age, were registered as voters with the Election Commission.
    \item 83 million new voters
    \item 15 million voters between age 18 and 19
\end{itemize}

\subsection{Polling}
Catering the general elections for around 850 million voters in India, the election was carried in 7 days (entire process will last 38 days) from April 11 to May 19, 2019. Some of the facts about the polling for General elections 2019 are:
\begin{table}
    \begin{center}

    \caption{2019 General Election Polling Facts}\label{tab1}
    \begin{tabular}{l|l}
    {\bf Item} &  {\bf Item Value} \\
    \hline
    Total number of polling booths &  1.035 million \\
    Maximum distance from voter's house &  2 kilometers \\
    Number of electronic voting machines & 3.96 million \\
    Election personnel being deployed & 11 million \\
    \end{tabular}
    \end{center}
\end{table}

\subsection{Expenditure}
Elections in India is definitely a costly affair as shown in Fig
\begin{figure}
    \centering
    \includegraphics[width=0.6\textwidth]{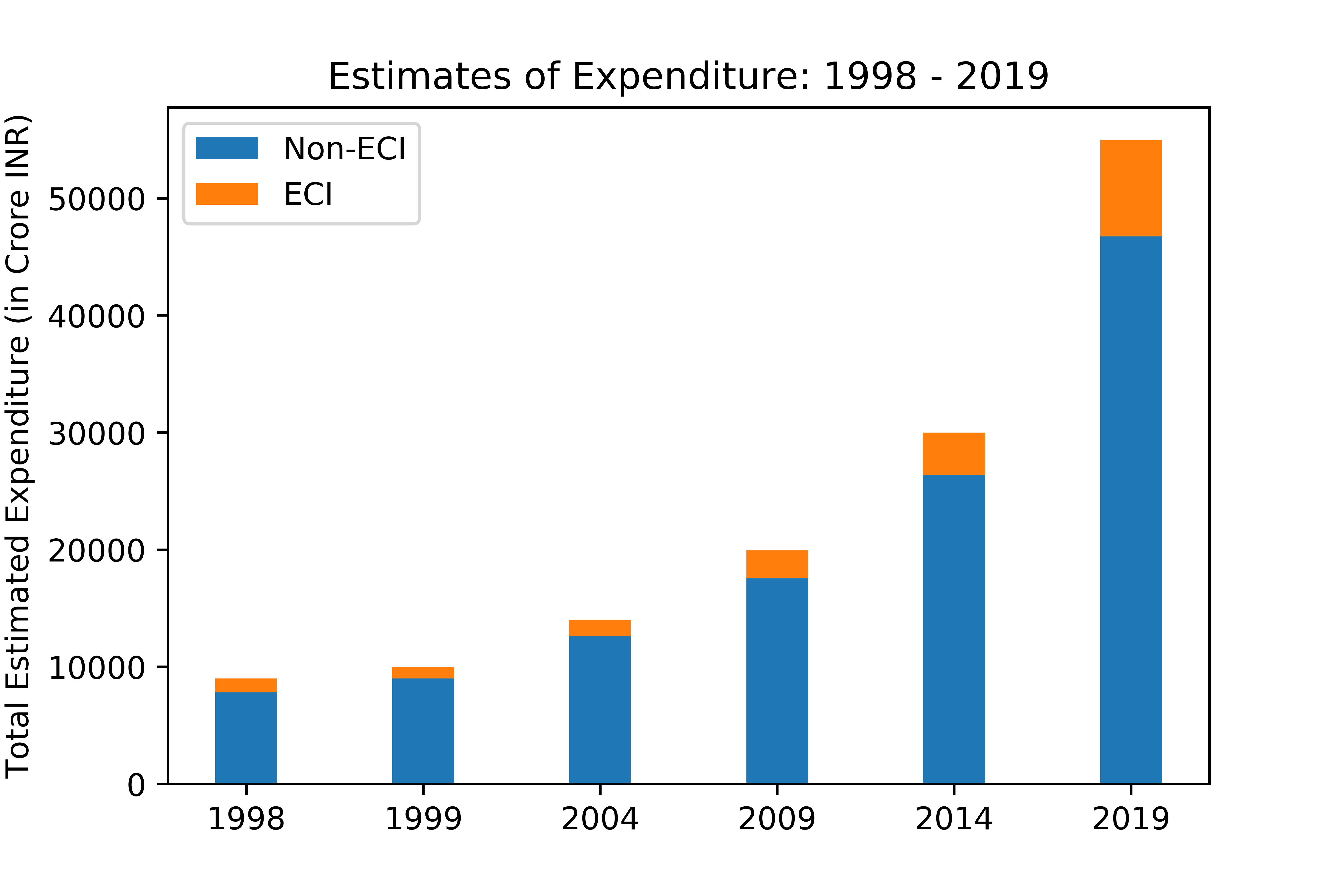}
    \caption{Estimated expenditure on Assembly elections in India. \cite{cms2019}} \label{fig1}
\end{figure}

\subsection{Electronic Voting Machines}

Electronic Voting Machine (also known as EVM) \cite{eci2019} is voting using electronic means to either aid or take care of the chores of casting and counting votes. Structure of the EVM is as follows: 
\begin{itemize}
    \item Units : Control and Balloting
    \item Connection : via a cable
    \item Control Unit : Polling Officer
    \item Polling officer verifies your identity and then enables ballot operation through the control unit. 
    \item Balloting Unit : Electors cast their votes through it. 
\end{itemize}

\textbf{Tampering in Electoral Voting Machines}
The Election Commission of India (ECI) has often faced flak due to allegations of millions of inaccurate names in voter rools, allowing code of conduct violations along with the biggest issue of malfuncting of the electronic voting machines. Following the silence of ECI on the same, also does not help with the credibility of EVMs, and they have been a constant subject of criticism till date. 

\section{National Financial Switch} \label{nfs}

National Financial Switch (NFS) is the largest network of shared automated teller machines (ATMs) in India. Institute for Development and Research in Banking Technology (IDRBT) started it in the year of 2004. Currently it is handled by National Payments Corporation of India (NPCI). It's main goal is to facilitate banking by connecting all the ATMs.

\begin{figure}
    \centering
    \includegraphics[width=\textwidth]{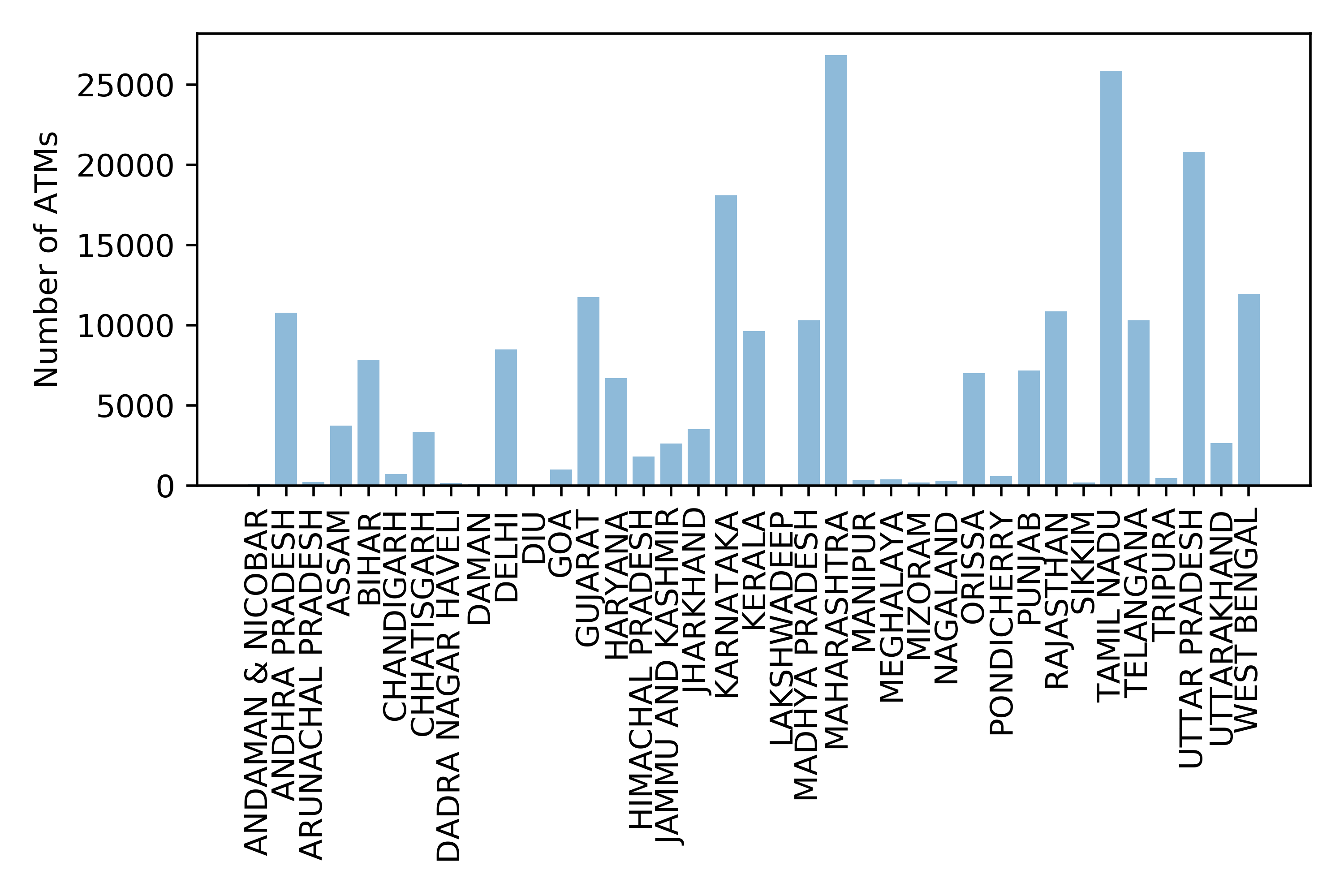}
    \caption{ATM distribution in India. \cite{RBI2019-1}} \label{atmd}
\end{figure}

According to the latest RBI statistics \cite{RBI2019}, there are about 4589727 POS machines and about 2,06,589 ATMs across India. The distribution of the same across Indian states is shown in the figure \ref{atmd}. The services rendered by the ATM machines are mini statements, change of PIN, enquiry of balance as well as cash withdrawal. 

The main features of the NFS ATMs and POS machines are secure, superb availability and instant transactions which are definitely one of the criteria that helps in choosing it the as the go to infrastructure for the polling for the elections. 

\section{Voting through NFS (ATMs + POS)} \label{proposal}

\begin{figure}
    \centering
    \includegraphics[width=\textwidth]{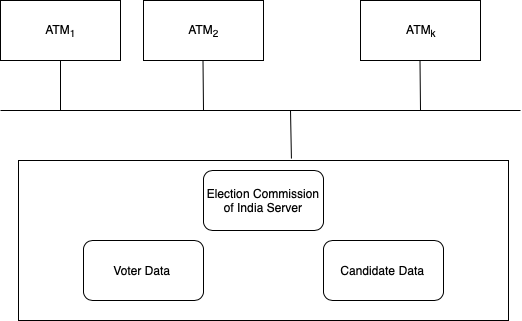}
    \caption{Architecture of the system proposed by the Google Patents \cite{Patent2003}} \label{prop1}
\end{figure}

Approach is to use the existing National Financial Switch architecture for secure, electronic and distributed voting through automated teller machines (ATMs) and Point of Sale Machines. The following points illustrate the overall architecture or methodology for the same. 
\begin{enumerate}
    \item Each voter is given unique voter identity card with photo on it that is ATM and POS readable.
    \item The same card should have the information about the voter and his constituency. This information is ofcourse encrypted on to the card. 
    \item All the information could be loaded on to the ATM machines/POS machines for the preparation of election day directly from the server. This process is equivalent to the current online testing that is conducted across India for various exams.
    \item Polling officer after verification of the card could allow the voter to enter into the ATMs to cast the vote.
    \item Each vote could be equivalent to a transaction on a credit/debit card and hence the security of a such transaction could be enhanced by the use of block chain.
    \item All voting or transaction will be securely documented and electronically verified. They will also be accompanied by the paper receipts and can also be accompanied by the photos of electors which would help in resolving discrepancy in case of their occurences. 
    \item Once the vote is cast, the voter would not be able to cast the vote at any other place. In case of multiple tries a flag could be raised to ECI for proper action.
\end{enumerate}

Tabulation of the above architecture in finding and tallying the votes for the candidate would follow the same secure channel and method to complete the process without any physical transport as in case of the EVMs. POS machines upon getting connectivity if they have missed the connectivity in difficult terrain could upload the number of votes, voter information and candidate information securedly to the ECI server.

\begin{figure}
    \centering
    \includegraphics[width=0.8\textwidth]{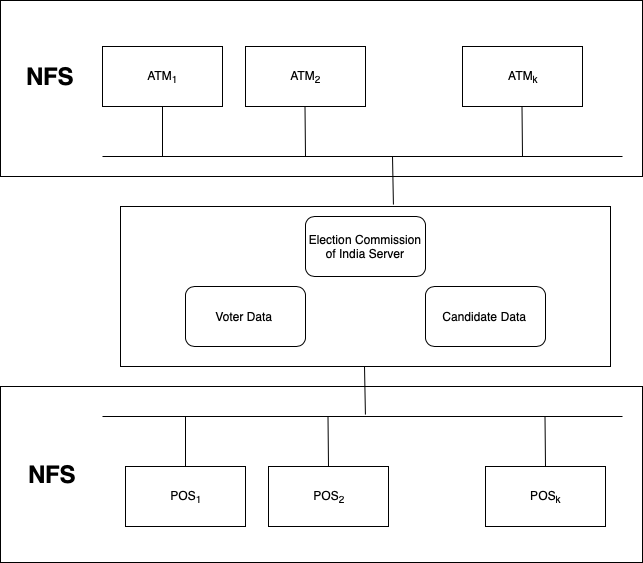}
    \caption{Proposed Architecture - II: This includes collaboration of both the Election Commission of India Server with the National Financial Switch. National Financial Switch comprises of the shared ATM network as well as POS machines.} \label{prop2}
\end{figure}
4
The advantage of an system such as this would also be extended to any location in the world (albeit by the coupling with the safe systems) for casting vote which is definitely a problem for Indian residents staying/working abroad but are willing to be a part of the democratic process of elections in India.

Other advantages of this system would be to use some already established systems such Aadhar IDs for the verification and setting up the initial voter indentity card. This would also help with adding the biometric security to enhance the credibility of the system to the criticism.

\section{Security Challenges}

There are many advantages to the use of ATM and POS machines through NFS for the election process in developing country in India. This could radically reduce the technological and financial overhead for such a grand process while adding transparency and fairness to the process.

However, the above system is still a target to threats given it lies in the intersection between the cyber and physical world. Minimal authentication, outdated operating system. The major attacks have been skimmer and cyber-attacks. Simply improving the physical security will become the significant deterrent to the attacks as most of them need physical access to the ATM and POS machines. Also the improvement in software systems such as block-chain and new encryption techniques along with realiable POS and ATM security monitoring and activity response service will enable Election Commission as well as citizens of our country to participate in this democratic exercise in a fair and transparent manner.

\section{Conclusion}

Indian elections serve as important mirror to the modern society about democratic process and enables millions of people to elect a government that work for their and country's interest. It is therefore imperative to have a fair election that also should be a burden on the tax payer. Hence our proposal involves using the existing National Financial Switch (NFS) infrastructure through the ATMs as well as POS machines to enable 850 million Indians to cast their vote. The new approach would definitely serve Indian people and Indian democracy in the right way. Future work would involve holding elections at a small scale in the university level through POS machines.

%
%
%
%

\end{document}